\documentclass[prd,preprint,nofootinbib]{revtex4-1} 

\usepackage{latexsym,verbatim}
\usepackage{amssymb}
\usepackage{amsfonts}
\usepackage{amsmath,color}
\usepackage{graphicx} 
\usepackage{bm}
\usepackage{physics}
\usepackage[utf8]{inputenc}
\usepackage[T1]{fontenc}

\usepackage{hyperref}

\hypersetup{
    colorlinks=true,
    linkcolor=red,
    citecolor=red,
    filecolor=magenta,      
    urlcolor=cyan,
    pdftitle={Quantum two-level systems and gravitational waves},
    pdfpagemode=FullScreen,}

\def\mb#1{\mathbf{#1}}

\def\ber{\begin{eqnarray}}
\def\eer{\end{eqnarray}}
\def\beq{\begin{equation}}
\def\eeq{\end{equation}}

\def\rmd{{\rm d}}

\def\ed{\end{document}}

\def\dT#1{\frac{\mathrm{d} #1}{\mathrm{d}T}}

\def\dTT#1{\frac{\mathrm{d} ^{2}#1}{\mathrm{d}T^{2}}}
\def\dt#1{\frac{\mathrm{d} #1}{\mathrm{d}t}}

\def\sT{\sin \left(\omega T \right)}
\def\cT{\cos \left(\omega T \right)}

\makeatletter
\let\jnl@style=\rm
\def\ref@jnl#1{{\jnl@style#1}}

\def\aj{\ref@jnl{AJ}}                   
\def\actaa{\ref@jnl{Acta Astron.}}      
\def\araa{\ref@jnl{ARA\&A}}             
\def\apj{\ref@jnl{ApJ}}                 
\def\apjl{\ref@jnl{ApJ}}                
\def\apjs{\ref@jnl{ApJS}}               
\def\ao{\ref@jnl{Appl.~Opt.}}           
\def\apss{\ref@jnl{Ap\&SS}}             
\def\aap{\ref@jnl{A\&A}}                
\def\aapr{\ref@jnl{A\&A~Rev.}}          
\def\aaps{\ref@jnl{A\&AS}}              
\def\azh{\ref@jnl{AZh}}                 
\def\baas{\ref@jnl{BAAS}}               
\def\bac{\ref@jnl{Bull. astr. Inst. Czechosl.}}
\def\caa{\ref@jnl{Chinese Astron. Astrophys.}}
\def\cjaa{\ref@jnl{Chinese J. Astron. Astrophys.}}
\def\icarus{\ref@jnl{Icarus}}           
\def\jcap{\ref@jnl{J. Cosmology Astropart. Phys.}}
\def\jrasc{\ref@jnl{JRASC}}             
\def\memras{\ref@jnl{MmRAS}}            
\def\mnras{\ref@jnl{MNRAS}}             
\def\na{\ref@jnl{New A}}                
\def\nar{\ref@jnl{New A Rev.}}          
\def\pra{\ref@jnl{Phys.~Rev.~A}}        
\def\prb{\ref@jnl{Phys.~Rev.~B}}        
\def\prc{\ref@jnl{Phys.~Rev.~C}}        
\def\prd{\ref@jnl{Phys.~Rev.~D}}        
\def\pre{\ref@jnl{Phys.~Rev.~E}}        
\def\prl{\ref@jnl{Phys.~Rev.~Lett.}}    
\def\pasa{\ref@jnl{PASA}}               
\def\pasp{\ref@jnl{PASP}}               
\def\pasj{\ref@jnl{PASJ}}               
\def\rmxaa{\ref@jnl{Rev. Mexicana Astron. Astrofis.}}%
\def\qjras{\ref@jnl{QJRAS}}             
\def\skytel{\ref@jnl{S\&T}}             
\def\solphys{\ref@jnl{Sol.~Phys.}}      
\def\sovast{\ref@jnl{Soviet~Ast.}}      
\def\ssr{\ref@jnl{Space~Sci.~Rev.}}     
\def\zap{\ref@jnl{ZAp}}                 
\def\nat{\ref@jnl{Nature}}              
\def\iaucirc{\ref@jnl{IAU~Circ.}}       
\def\aplett{\ref@jnl{Astrophys.~Lett.}} 
\def\apspr{\ref@jnl{Astrophys.~Space~Phys.~Res.}}
\def\bain{\ref@jnl{Bull.~Astron.~Inst.~Netherlands}}
\def\fcp{\ref@jnl{Fund.~Cosmic~Phys.}}  
\def\gca{\ref@jnl{Geochim.~Cosmochim.~Acta}}   
\def\grl{\ref@jnl{Geophys.~Res.~Lett.}} 
\def\jcp{\ref@jnl{J.~Chem.~Phys.}}      
\def\jgr{\ref@jnl{J.~Geophys.~Res.}}    
\def\jqsrt{\ref@jnl{J.~Quant.~Spec.~Radiat.~Transf.}}
\def\memsai{\ref@jnl{Mem.~Soc.~Astron.~Italiana}}
\def\nphysa{\ref@jnl{Nucl.~Phys.~A}}   
\def\physrep{\ref@jnl{Phys.~Rep.}}   
\def\physscr{\ref@jnl{Phys.~Scr}}   
\def\planss{\ref@jnl{Planet.~Space~Sci.}}   
\def\procspie{\ref@jnl{Proc.~SPIE}}   

\makeatother

\begin{document}

\author{Matteo Luca Ruggiero}
\email{matteoluca.ruggiero@unito.it}
\affiliation{Dipartimento di Matematica ``G.Peano'', Universit\`a degli studi di Torino, Via Carlo Alberto 10, 10123 Torino, Italy}
\affiliation{INFN - LNL , Viale dell'Universit\`a 2, 35020 Legnaro (PD), Italy}

\date{\today}

\title{Quantum two-level systems and gravitational waves}

\begin{abstract}
We study the interaction between gravitational waves and a quantum two-level system consisting of a spin 1/2  particle using the formalism of the proper detector frame. This approach highlights the effects of gravitational waves on both the particles and the observer, emphasizing that only relative measurements can be made. Specifically, within this framework, the gravitational field of the waves is described using the gravitoelectromagnetic analogy. The interaction of the  system is then determined by the gravitomagnetic field of the wave, which induces a time-dependent perturbation. We analyze this perturbation for both generic frequencies and resonance conditions, and discuss its implications.
\end{abstract}

\maketitle

\section{Introduction} \label{sec:intro}

The best models we have to describe physical phenomena in the Universe, Quantum Mechanics (QM) and General Relativity (GR), are based on two seemingly conflicting hypotheses. QM posits that the ultimate nature of matter and fields is discrete, while GR describes gravitational interactions as the geometrical structure of the spacetime continuum.  In Einstein's words: ``The problem seems to me how one can formulate statements about a discontinuum without calling upon a continuum (spacetime) as an aid''  (quoted by \citet{stachel1986einstein}) and, even today, the relationship between these two models, with the ultimate goal to reconcile them within a common paradigm, remains one of most challenging taks in theoretical physics.  As a matter of fact, a semiclassical approach, where gravity is treated as a classical field, successfully describes all known phenomena involving the interplay between gravitational and quantum fields: in other words, as suggested by \citet{Parikh:2020fhy}, we have hardly any evidence at all that gravity is quantized. Remarkably, a recent proposal by \citet{oppenheim23} introduces a consistent theory of classical gravity coupled to quantum field theory. 

Since their first detection \cite{PhysRevLett.116.061102}, gravitational waves (GWs) have become a powerful tool for exploring the Universe, complementing electromagnetic radiation. Additionally, they present an important opportunity to study the potential consequences of the quantization of gravity using suitably designed detectors. In this context, it is interesting to consider the interaction between GWs, treated as classical fields, and matter, treated  with a quantum mechanical approach. A systematic study of this problem was outlined by \citet{spelio}, who considered the interaction between GWs and   quantum mechanical systems, under the long wavelength approximation for the gravitational field of the waves and the non relativistic limit for the systems considered, which are the free particle, the two-dimensional harmonic oscillator and the hydrogen atom. The approach adopted is based on the canonical quantization procedure, starting from the interaction Hamiltonian. As subsequently pointed out by \citet{spelio2}, in this scenario it is very relevant to correctly describe the measurement process, due to the peculiarity of the gravitational interaction. The latter, unlike other fundamental interactions, cannot be screened. As a consequence, it is necessary to include the effects of the GWs on the observer or measuring device to properly describe the result of an experiment. To fix the ideas, let us consider the description of the effect of GWs on a free particle, as seen by a geodesic observer. Both the observer and the particle are moving along their geodesic worldlines: the passage of GWs provokes tidal forces, which change the \textit{relative position and velocity} of the free particle with respect to the observer. The key point is that the observer cannot measure the motion of the particle independently of their motion in the field of the GWs. Consequently, it is reasonable to expect that any measurable effects due to the passage of GWs must account also for the effect of the GWs on the measuring device. This process can be naturally formalized in the \textit{proper detector frame or laboratory frame}, which serves as an alternative to the transverse-traceless (TT) frame \cite{maggiore2007gravitational,misner} for describing the effects of gravitational waves. In particular, the spacetime metric in the proper detector frame is based on the use of {Fermi coordinates}, which  are a quasi-Cartesian coordinates system that can be build in the neighbourhood of the world-line of a reference observer, and their definition depends both on the background field where the observer is moving and, also, on the kind of motion. Fermi coordinates are  defined, by construction, as scalar invariants \cite{synge1960relativity}; they have a concrete meaning, since they are the coordinates an observer would naturally use to make space and time measurements in the vicinity of their world-line.  In the laboratory frame, all measurements are expressed relative to the reference observer \cite{Ruggiero:2021qnu}. 

{Recent studies have explored the interaction of gravitational waves (GWs) with quantum systems. For instance, \citet{gwquantum1} consider a detector consisting of two uncoupled one-dimensional anisotropic oscillators, demonstrating that GWs can be detected through the induced Berry phase. Similarly, \citet{gwquantum2} investigate quantum entanglement between two mutually orthogonal modes of simple harmonic oscillators, suggesting to use LIGO’s arms as oscillators. In this context, it is important to note that experiments involving gravitational waves \cite{gwq2,gwq3} are potential candidates for probing the nature of gravity—whether it behaves as a classical or a quantum field. One approach to addressing this question is to design an experimental setup that can be analyzed under both classical and quantum gravity frameworks, yielding distinct predictions in each case \cite{gwq1}. Although our approach treats the gravitational field classically, these studies place our work within a broader research framework.}

The purpose of this paper is to focus on the interaction between GWs   and a quantum system consisting of a spin 1/2 particle, described as a two-level system. {A two-level system has been used as a quantum clock model to study interferometric effects in atoms propagating through a stationary gravitational field generated by Newtonian sources \cite{Roura:2018cfg}. To the best of our knowledge, our work is the first to consider the interaction of a two-level system with GWs.}

 In a previous publication \cite{Ruggiero_2020b} we explored the interaction of GWs with a spinning particle using the gravitoelectromagnetic analogy \cite{Ruggiero:2023ker} arising in the laboratory frame \cite{Ruggiero_2020}. Specifically, we highlighted that the interaction between GWs and spin gives rise to what we termed gravitomagnetic resonance. We initially described this effect for a classical spinning particle and then provided a heuristic description for a quantum spinning particle based on the analogy with the corresponding electromagnetic case. {The classical interaction between GWs and spinning particle was studied by \citet{biniortolan2017}.} Here, we use the standard approach for time-dependent perturbations in quantum mechanics to describe the interaction of a spin 1/2 particle with a gravitational wave. We consider both the general case of perturbation and the resonant one.

The paper is organised as follows: in Section \ref{sec:fermi} we recall the essential features of the spacetime metric in the Fermi frame, utilizing  the gravitoelectromagnetic analogy. Then, in Section \ref{sec:tdp} we study the effects of GWs trated as a time-depending perturbation on a two-level system, while in  Section \ref{sec:secres} we discuss perturbations that are in resonance with the system's proper frequency. Finally, we present our discussion  and conclusions  in Section \ref{sec:disconc}.

\section{The Metric in the Fermi Frame} \label{sec:fermi}

{In General Relativity, coordinates are merely labels used to distinguish spacetime events and, generally speaking, are not directly related to physically measurable quantities. Of course, measurable effects are independent of the method used to parameterize spacetime events, and the choice of coordinates is often guided by descriptive simplicity and the symmetries of the problem under consideration. However, when measurements are performed in a laboratory, it is important to establish a connection between the actual rulers and clocks used by the observer—who locally perceives a Minkowskian spacetime—and the background spacetime, which encodes gravitational effects. In particular, \textit{Riemann coordinates} provide the closest approximation to a Minkowski frame that an observer can construct in curved spacetime near a specific spacetime event. However, since these coordinates are tied to a particular spacetime point, their construction must be repeated to connect measurements taken at different events. On the other hand, \textit{Fermi (normal) coordinates} offer the closest approximation to a Cartesian frame that an observer can construct in curved spacetime along their world-line. Thus, they are particularly well-suited for describing measurement processes in a straightforward manner. A thorough discussion of different types of coordinate systems in the context of gravitational waves can be found in the work of \citet{Rakhmanov_2014}.}

So, the spacetime metric in the laboratory frame is obtained using Fermi coordinates, which are constructed as scalar invariants and, hence, allow a natural definition of the measurement process in terms of observable quantities. This metric depends both on how the laboratory frame moves in spacetime (i.e. on its acceleration and the rotation of its axes) and on the spacetime curvature through the Riemann  tensor \citep{Ni:1978di,Li:1979bz,1982NCimB..71...37F,marzlin}. In particular, exploiting the gravitoelectromagnetic analogy discussed by  \citet{Ruggiero_2020}, using coordinates $(cT,X,Y,Z)$, for a laboratory in free-fall and non rotating, up to quadratic displacements $|X^{i}|$ from the reference world-line, the metric can be written in the form\footnote{Latin indices refer to space coordinates, while Greek indices to spacetime ones. Moreover,  we will use bold-face symbols like  $\mb W$ to refer to vectors in the Fermi frame. The spacetime signature is $(-1,1,1,1)$.}
\beq
\mathrm{d} s^2= -c^2 \left(1-2\frac{\Phi}{c^2}\right)\rmd T^2 -\frac4c \mathcal A_{i}\rmd X^{i}\rmd T  +
 \left(\delta_{ij}+2\frac{\Psi_{ij}}{c^2}\right)\rmd X^i \rmd X^j\ , \label{eq:weakfieldmetric11}
\eeq
with
\begin{eqnarray}
\Phi (T, { X^{i}})&=&-\frac{c^{2}}{2}R_{0i0j}(T )X^iX^j, \label{eq:defPhiG}\\
\mathcal A^{}_{i}(T ,{X^{i}})&=&\frac{c^{2}}{3}R_{0jik}(T )X^jX^k, \label{eq:defAG}\\
\Psi_{ij} (T, {X^{i}}) & = & -\frac{c^{2}}{6}R_{ikjl}(T)X^{k}X^{l}, \label{eq:defPsiG}
\end{eqnarray}
where $\Phi$ and $\mathcal A_{i}$ are, respectively, the {gravitoelectric} and {gravitomagnetic} potential, and $\Psi_{ij}$ is the perturbation of the spatial metric \cite{manasse1963fermi,MTW,Mashhoon:2003ax,Ruggiero_2020,Ruggiero:2021uag,Ruggiero:2023ker}. In the above equations $R_{\alpha \beta \gamma \delta}(T)$ is the projection of the Riemann tensor on the tetrad attached to the laboratory frame, and it is evaluated  along the reference world-line. The gravitoelectromagnetic analogy allows to express the geodesics equation in terms of Lorentz-force: in other words, the motion of test particles relative to the reference observer is determined by the action of (tidal) gravitoelectromagnetic fields, that can be obtained from the corresponding potentials   \cite{Ruggiero:2021uag}.  As for us, we are interested in the laboratory metric describing the field of a plane gravitational wave propagating in the $X$ direction with frequency $\omega$.  In this case, the components of the gravitomagnetic field  $\displaystyle \mb B= \bm \nabla \wedge \mb{\mathcal A}$  turn out to be {(see Appendix \ref{sec:app})}:
\begin{eqnarray}
B^{}_{X}(T)  &=& 0, \label{eq:BX} \\
 B^{}_{Y}(T)  &=& -\frac{\omega^{2}}{2}\left[-A^{\times} \cT Y+A^{+} \sT Z \right]  \label{eq:BY}, \\
  B^{}_{Z}(T)  &=& -\frac{\omega^{2}}{2}\left[A^{+}\sT Y+A^{\times}\cT Z \right] \label{eq:BZ}.
\end{eqnarray}
where  $A^{+},A^{\times}$ are the amplitude of wave in the two polarization states. We point out that the expression of the metric tensor used to obtain the gravitomagnetic field is valid in the large wavelength limit, which means that the typical dimension $\ell$ of the laboratory frame  is negligible with respect to the wavelength $\lambda$. More accurate expressions can be obtained, which contain higher order terms in the small parameter $\displaystyle \epsilon = \frac{\ell}{\lambda}$  (see e.g. \citet{Ruggiero:2022gzl} and references therein).  The series expansion can be exactly summed to obtain a compact form \cite{1982NCimB..71...37F,Berlin:2021txa}.

 Actually, if the laboratory is rotating there is an additional contribution to the gravitomagnetic field which depends on the rotation rate of the laboratory frame with respect to non rotating Fermi-Walker one. In particular, this contribution is in the form
\beq
\mb B_{0}= -\mb \Omega c, \label{eq:defB0}
\eeq
where $\mb \Omega$ is the rotation rate \cite{Ruggiero_2020}. Here we suppose that $\mb B_{0}=-\Omega c \mb u_{X}$, i.e. that the laboratory is rotating around the direction of propagation of the wave.  In summary,  in the laboratory frame we have a total gravitomagnetic field in the form
\beq
\mb B_{\mathrm{tot}}(T)= \mb B_{0}+\mb B(T)=B_{0}\mb u_{X}+B_{Y}(T)\mb u_{Y}+B_{Z}(T)\mb u_{Z},
\eeq
with respect to a basis of unit vectors $\left(\mb u_{X}, \mb u_{Y}, \mb u_{Z}\right)$. We remark that $\mb B_{0}$ is a constant field while the gravitomagnetic field determined by the passage of the wave is time-depending according to Eqs. (\ref{eq:BY})-(\ref{eq:BZ}). 

{Based on the discussion presented by \citet{mashhoon_entropy}, we assume that, in the correspondence limit, a particle with intrinsic spin behaves like an ideal gyroscope. This assumption is rooted in the well-known gravitomagnetic analogy \cite{Mashhoon:2003ax}. Consequently, the interaction between the spin $\mb S$  and the inertial and gravitational effects in the laboratory frame can be described using the Hamiltonian
\beq
H(T)=\frac 1 c \mb S \cdot \mb B_{\mathrm{tot}}(T)=\frac 1 c \mb B_{0} \cdot \mb S+\frac 1 c \mb B_{}(T) \cdot \mb S. \label{eq:Hcoupling}
\eeq
In particular, this result can also be derived from the Dirac equation in a Fermi frame, as demonstrated by \citet{HehlPhysRevD.42.2045}. A detailed discussion is provided in the work of \citet{SPINMashhoon:1997qc}. In addition, we emphasize that our approach focuses solely on the spin degrees of freedom. This implies that the external degrees of freedom, such as position and momentum, can be neglected, similar to the case of the Stern-Gerlach experiment.}

In what follows, the term
\beq
H_{0}=\frac 1 c \mb B_{0} \cdot \mb S \label{eq:H0def}
\eeq
will be referred to as the unperturbed Hamiltonian, meanwhile
\beq
W(T)=\frac 1 c \mb B_{}(T) \cdot \mb S \label{eq:Wtpert}
\eeq
will be considered the perturbation.

\section{Time-depending Perturbations} \label{sec:tdp}

Our quantum system consists of a spin 1/2 particle, which can be treated as a two-level system. 

The general analysis proceeds as follows: we suppose that the system  is initially in the stationary state $\ket \phi_{i}$ of the Hamiltonian $H_{0}$; for the latter operator we have $\displaystyle  H_{0}\ket{\phi_{n}}=E_{n}\ket{\phi_{n}}$, where $\{\ket{\phi_{n}}\}$  is a complete set of  eigenvectors corresponding to the eigenvalues $E_{n}$. Then, at $T=0$ a perturbation $W(T)$ starts acting on the system, so that we have 
\beq
 H(T)=H_{0}+ W(T), \label{eq:Ht}
\eeq
with $W(T)=\sigma \hat W(T)$, where $\sigma \ll 1$ is a real dimensionless number, while $\hat W(T)$ is of the same order as $H_{0}$. Generally, if we suppose that at $T=0$ the system is in the eigenstate $\ket{ \phi_{i}}$ of $H_{0}$, for $T>0$ this will not be an eigenstate of $ H(T)$. What we are interested in is the probability of finding the system in a different eigenstate $\ket{\phi_{f}}$ of $ H_{0}$ at the generic time $T>0$.  In other words, we assume that the perturbation causes the system to evolve towards the eigenstates of the unperturbed Hamiltonian $H_{0}$. 

In our case and referring to Eq. (\ref{eq:Hcoupling}), we suppose that the system for $T<0$ is in an eigenstate of $\displaystyle H_{0}=\frac 1 c B_{0}S_{X}$, and it is perturbed by $\displaystyle W(T)=\frac 1 c\left[B_{Y}(T)S_{Y}+B_{Z}(T)S_{Z} \right]$.  

We remark that $W(T)$ is proportional to the gravitomagnetic field: since for GWs reaching the Earth we have  $A^{+} \simeq 10^{-21 } \simeq A^{\times}$, we see that it can be safely considered as a perturbation.
In particular, 
taking into account the expression of the gravitomagnetic field given in Eqs. (\ref{eq:BX} )-(\ref{eq:BZ}), we obtain
\beq
W(T)=-\frac{\omega^{2}}{2c}\left[A^{\times}\cT \left(-S_{Y}Y+S_{Z}Z \right)+A^{+}\sT \left(S_{Y}Z+S_{Z}Y \right) \right]. \label{eq:defWt}
\eeq 
The standard approach for time-depending perturbations (see e.g. \citet{cohen1991quantum}) can be applied. To this end, for the system under consideration we choose a complete set of observable made of the eigenstates of $ S_{X}$, $\{\ket{+},\ket{-} \}$, so that
\beq
 S_{X}\ket{+}=\frac{\hbar}{2}\ket{+}, \quad  S_{X}\ket{-}=-\frac{\hbar}{2}\ket{-}.  \label{eq:SX}
\eeq
Accordingly, the representation of the three operators $ S_{X},  S_{Y},  S_{Z}$ is 
\beq
 S_{X}= \frac{\hbar}{2} \left(\begin{array}{cc}1 & 0 \\0 & -1\end{array}\right), \quad S_{Y}= \frac{\hbar}{2} \left(\begin{array}{cc}0 & 1 \\1 & 0\end{array}\right), \quad S_{Z}= \frac{\hbar}{2} \left(\begin{array}{cc}0 & -i \\i & 0\end{array}\right). \label{eq:defSpauli}
\eeq

For the sake of simplicity and to evaluate the impact of the effect,  we set $A^{+}=0$ in the perturbation term (\ref{eq:defWt}), and we get
\beq
W(T)=W^{\times}\cT  \left(\begin{array}{cc}0 & e^{i\theta^{\times}} \\ e^{-i\theta^{\times}} & 0\end{array}\right), \label{eq:defWt1}
\eeq
where $\displaystyle W^{\times}=\frac{\omega^{2}}{4c}A^{\times} k_{0}{\hbar}$, $\displaystyle k_{0}=\sqrt{Y^{2}+Z^{2}}$, and $\tan \theta^{\times}=\frac{Z}{Y}$.
Up to first order in the perturbation, the transition probability is given by
\beq
P_{if}=\frac{1}{\hbar^{2}} \left| \int_{0}^{t}e^{i\omega_{fi}t'}W_{fi}(t')dt'  \right|^{2}, \label{eq:solpert30}
\eeq
where 
\beq
\mel{\phi_{f}}{ W}{\phi_{i}}=W_{fi} \label{eq:Sch4}
\eeq
are the matrix elements of the perturbation $ W(t)$, and $\omega_{fi}$ are the frequencies related to the energy difference between the two states:
\beq
\omega_{fi}=\frac{E_{f}-E_{i}}{\hbar}. \label{eq:Sch9}
\eeq
Accordingly the matrix elements are
\beq
W_{12}=W^{\times}\cT e^{i\theta^{\times}}, \quad W_{21}=W^{*}_{12},  \label{eq:defWt2}
\eeq
We can calculate the energy difference between the two eigenstates $\ket{+},\ket{-} $:
\begin{eqnarray}
H_{0}\ket{+}&=& \frac 1 c B_{0} \frac{\hbar}{2} \ket{+}=E_{1}\ket{+}, \label{eq:eigen1}\\
H_{0}\ket{-}&=& -\frac 1 c B_{0} \frac{\hbar}{2} \ket{-}=E_{2}\ket{+}. \label{eq:eigen2}
\end{eqnarray}
Consequently,  from Eq. (\ref{eq:Sch9}) we obtain
\beq
\omega_{12}=\frac 1 c B_{0}. \label{eq:omega120}
\eeq

If we calculate the transition probability (\ref{eq:solpert30}) from the initial state $\ket{+}$ to $\ket{-}$ we get \beq
P_{12}=\frac{\left|{W^{\times}}\right|^{2}}{4\hbar^{2}}\left| \frac{1-e^{i\left(\omega_{21}+\omega \right)T}}{\omega_{21}+\omega}+\frac{1-e^{i\left(\omega_{21}-\omega \right)T}}{\omega_{21}-\omega} \right|^{2}. \label{eq:transquad12}
\eeq

We see that $P_{12}$  has a maximum, at fixed time, for $\omega=\pm \omega_{21}$: for instance, when  $\omega_{21}-\omega \ll \omega_{21}$, we can neglect the first term in  (\ref{eq:transquad12}) and we have $\displaystyle P_{12}=\frac{\left|{W}^{\times}\right|^{2}}{4\hbar^{2}}\left[\frac{\sin \left(\omega_{21}-\omega \right)T/2}{ \left(\omega_{21}-\omega \right)/2}\right]^{2}$.  The same can be done when $\omega_{21}+\omega \ll \omega_{21}$. 

In general, we see that the transition probability is a \textit{quadratic} function of the GWs amplitude: although our analysis was limited  to the $A^{\times}$ polarization, the same results apply to  arbitrary polarization of the wave. Regardless, this interaction remains exceedingly small for gravitational waves reaching the Earth.

That being said, we point out that at the maximum $P_{12}=\frac{\left|{W}^{\times}\right|^{2}}{4\hbar^{2}} T^{2}$: accordingly,  this method cannot be used for arbitrary large times, since it should always be $P_{12} \leq 1$.  Therefore, under resonance conditions  where $\omega_{12}=\pm \omega$  we need to adopt a different approach. As we will discuss below, the effect of the perturbation on the system in this scenario differs, and the role of the perturbation amplitude changes.

\section{Secular Effects of a Resonant Perturbation} \label{sec:secres}

Now, we focus on the effect of a resonant perturbation. To begin with, we note from Eqs. (\ref{eq:defB0}) and  (\ref{eq:omega120}) that the proper frequency of the system is given by the rotation rate of the laboratory frame. So, we consider the situation when $\omega=\Omega$. We recall that the evolution of the system is determined by the Schr\"odinger equation
\beq
i\hbar \dT{\ket{\psi(T)}}=\left[ H_{0}+{W(T)} \right] \ket{\psi(t)}, \label{eq:Sch1}
\eeq
with the initial condition
\beq
\ket{\psi(t=0)}=\ket{\phi_{i}}. \label{eq:Sch2}
\eeq
The state vector can be generally written as
\beq
\ket{\psi(T)}=\sum_{n}c_{n}(T)\ket{\phi_{n}}, \label{eq:Sch3}
\eeq
where $\displaystyle c_{n}(T)=\bra{\phi_{n}}\ket{\psi(T)}$. On setting 
\beq
c_{n}(t)=b_{n}(t)e^{-\frac{iE_{n}t}{\hbar}}, \label{eq:Sch8}
\eeq
we obtain
\beq
i\hbar \dt{b_{n}(t)}=\sum_{k}  W_{nk} b_{k}(t) e^{i \omega_{nk}t}, \label{eq:Sch10}
\eeq
which is equivalent to the  Schr\"odinger equation (\ref{eq:Sch1}). In addition, we notice that since the 
transition probability is $\displaystyle {P}_{if}=\left|\bra{\phi_{f}}\ket{\psi(t)} \right|^{2}$, taking into account Eqs. (\ref{eq:Sch3}) and (\ref{eq:Sch8}) we may write
\beq
P_{if}=|c_{f}(t)|^{2}=|b_{f}(t)|^{2}. \label{eq:probtrans2}
\eeq
For a two-level system Eq. (\ref{eq:Sch10}) boils down to:
\begin{eqnarray}
i\hbar\dT{b_{1}(T)}&=&W_{11} b_{1}(T)+W_{12}b_{2}(T)e^{i\omega_{12}T}, \label{eq:schres11}\\
i\hbar\dT{b_{2}(T)}&=&W_{21} b_{1}(T)e^{i\omega_{21}T}+W_{22}b_{2}(T). \label{eq:schres12}					
\end{eqnarray}

As we have seen, in the case of a plane gravitational wave, the perturbation is in the general form (\ref{eq:defWt}): in particular, we have two terms whose time-dependence is defined in terms of trigonometric functions $\cT,\sT$. Let us consider for simplicity the case $A^{\times}=0$, so that
\beq
W(T)=W^{+} \sT  \left(\begin{array}{cc}0 & e^{-i\theta^{+}} \\ e^{i\theta^{+}} & 0\end{array}\right), \label{eq:defWplus}
\eeq
where $\displaystyle W^{+}=-\frac{\omega^{2}}{4c}A^{+} k_{0}{\hbar}$,  $\displaystyle k_{0}=\sqrt{Y^{2}+Z^{2}}$, and $\tan \theta^{+}=\frac{Y}{Z}$.
Then, Eqs. (\ref{eq:schres11})-(\ref{eq:schres12}) become
\begin{eqnarray}
i\hbar\dT{b_{1}(T)}&=&\frac{1}{2i}\left[{\overline{W}_{12}}b_{2}(T) \left(e^{i\left(\omega-\omega_{21} \right)T}-e^{i\left(\omega+\omega_{21} \right)T} \right) \right] \label{eq:schres21}, \\
i\hbar\dT{b_{2}(T)}&=&\frac{1}{2i}\left[{\overline{W}_{21}}b_{1}(T) \left(e^{i\left(\omega+\omega_{21} \right)T}-e^{i\left(\omega_{21}-\omega \right)T} \right) \right],
\end{eqnarray}
where
\beq
\overline{W}_{12}=W^{+} e^{-i\theta^{+}}, \quad \overline{W}_{21}=\overline{W}^{*}_{12}.  \label{eq:defWt2sec}
\eeq


If we are near the resonant condition $\omega_{21}\simeq\omega$, we see that all terms are rapidly oscillating except those for which the exponential functions are almost constant. So, in this condition, we may write
\begin{eqnarray}
i\hbar\dT{b_{1}(T)}&=&\frac{1}{2i}{\overline{W}_{12}}b_{2}(T) e^{i\left(\omega-\omega_{21} \right)T},   \label{eq:schres31} \\
i\hbar\dT{b_{2}(T)}&=&-\frac{1}{2i}{\overline{W}_{21}}b_{1}(T) e^{i\left(\omega_{21}-\omega \right)T}.   \label{eq:schres32} 
\end{eqnarray}
If we set $\omega_{21}=\omega$, the above system can be exactly solved. In fact, it leads to
\beq
\dTT{b_{1}(T)}+\frac 1 4 \frac{|{\overline{W}_{21}|^{2}}}{\hbar^{2}}b_{1}(T)=0, \label{eq:osci1}
\eeq
which is the equation of a harmonic oscillator. On setting the initial conditions
\beq
b_{1}(0)=1, \quad b_{2}(0)=0, \quad \dT{b_{1}(0)}=0, \dT{b_{2}(0)}=\frac{1}{2\hbar} {\overline{W}_{21}}, \label{eq:osci2}
\eeq
we get 
\beq
b_{1}(T)=\cos\left(\frac 1 2 \frac{|\overline{W}_{21}|}{\hbar} T \right), \quad b_{2}(T)= e^{i\theta^{+}}\sin\left(\frac 1 2 \frac{|\overline{W}_{21}|}{\hbar} T \right), \label{eq:osci3}
\eeq
As a consequence, the  transition probability is
\beq
P_{12}=\sin^{2}\left(\frac 1 2 \frac{|\overline{W}_{21}|}{\hbar} T \right), \label{eq:osci4}
\eeq 
and we see that this probability depends on the amplitude of the gravitational wave through $|\overline{W}_{21}|=|W^{+}|$.  We see that, \textit{independently of} the amplitude of the GWs, the transition probability reaches $1$  for the first time when
\beq
T=\pi \frac{\hbar}{\overline{W}_{12}}. \label{eq:defTtrans}
\eeq
Accordingly, the GWs amplitude has the role to define the time required for the transition from one state to the other, but it does not affect the  transition probability as in the case of non resonant perturbations.
We point out that the same result is obtained for a perturbation in the form (\ref{eq:defWt2}), or, more generally speaking, with a perturbation in the form (\ref{eq:defWt1}).

 We conclude by adding that if $\omega \simeq \omega_{21}$ the system of equation (\ref{eq:schres31})-(\ref{eq:schres32}) leads to the solution given by the Rabi's formula
\beq
P_{12}(T)=\frac{|W_{12}|}{|W_{12}|^{2}+\hbar^{2}\Delta \omega^{2}} \sin^{2} \left (\sqrt{\frac{|W_{12}|^{2}}{\hbar^{2}}+\Delta \omega^{2}} \frac T 2 \right) \label{ee:rabi1}
\eeq
where $\Delta \omega=\omega-\omega_{21}$.

\section{Discussion and Conclusions} \label{sec:disconc}

We studied the interaction between a quantum system consisting of a spin 1/2  particle and the field of a plane gravitational wave. To do this, we examined the interaction in the laboratory frame, where the effects of the gravitational waves  can be described in terms of tidal gravitoelectromagnetic fields. This approach highlights that the measurement process involves both the observer and the system, as the observer cannot describe the effects on the particles independently of their own motion in the GWs field. In other words, measurements are relative to those made by the reference observer: the perturbation term in Eq. (\ref{eq:defWt})  is zero along the reference world-line, indicating that the evolution of spinning particles is relative to those located along this world-line, which remain unperturbed.

In the interaction with the GWs, we focused on the spin degrees of freedom and neglected the external position and momentum variables, which means that they can be treated classically or are not relevant for the measurement process.

 In this context, the interaction is described by the Hamiltonian (\ref{eq:Hcoupling}), which highlights the role of the gravitomagnetic field in the laboratory frame: in particular, before the passage of the wave the particles are in the eigenstates of the unperturbed Hamiltonian (\ref{eq:H0def}), which are determined by the rotation of the reference frame.  Due to the passage of the wave, we have an additional component of the gravitomagnetic field given by (\ref{eq:BX})-(\ref{eq:BZ}). In particular, this additional component can be safely considered as a perturbation, since the amplitude of the GWs reaching the Earth is very small. Accordingly, the whole process can be described in terms  of time-depending perturbations of two-level systems.

Our objective was to evaluate the transition probability  induced by the passage of GWs. First, we considered GWs with arbitrary frequency. Our analysis reveals that this transition probability is a quadratic function of the wave amplitude. Therefore, in realistic scenarios where GWs are detectable on Earth, we expect this probability to be very small.

The situation differs for resonant perturbations, which in our case correspond to gravitational waves with a frequency matching that of the rotating frame. We demonstrated that, regardless of the amplitude of the gravitational waves, when the resonance condition is met, the transition probability reaches 1 within the time interval specified in  Eq. (\ref{eq:defTtrans}). In a previous work \cite{Ruggiero_2020b} we arrived at similar results utilizing a heuristic argument for circularly polarized gravitational waves, and referred to this effect as gravitomagnetic resonance.  Our findings here show that this effect is quite general. An estimate of the order of magnitude of time interval (\ref{eq:defTtrans})  can be obtained since $\displaystyle T \simeq \frac{1}{\frac{\ell}{\lambda} A\,\omega}$, where $A=|A^{+}|\simeq |A^{\times}|$, and we must remember that we are working in the long-wavelength limit, hence $\displaystyle \frac{\ell}{\lambda} \ll 1$. In realistic situations, achieving arbitrarily high rotation frequencies is not feasible. For macroscopic systems, an upper limit of around $10^{3}$  Hz can be set. When combined with the amplitude of gravitational waves reaching the Earth, this suggests that the transition time is very long. However, as we proposed \cite{Ruggiero_2020b}, considering charged spinning particles could offer an equivalent scenario by utilizing a true magnetic field. According to the Larmor theorem, there is an equivalence between a system of electric charges in a magnetic field and the same system rotating at the Larmor frequency.

As previously mentioned, our approach considered only the spin degrees of freedom, excluding position and momentum, similar to the Stern-Gerlach experiment. In this context, it is noteworthy that, as suggested by \citet{mashhoon_entropy}, a gravitomagnetic Stern-Gerlach force arises. It  can be derived from the interaction Hamiltonian (\ref{eq:Hcoupling}) and its components are $f_{i}=-\partial_{i} H$: in particular, it is entirely determined by the non uniform gravitomagnetic field (\ref{eq:BX})-(\ref{eq:BZ}). For instance, if the spinning particles are prepared with $S_{Z}=0$, the components of the Stern-Gerlach force turn out to be $\displaystyle f_{Y}=-\frac{\omega^{2}}{2c}S_{Y}A^{\times}\cT, f_{Z}=\frac{\omega^{2}}{2c}S_{Y}A^{+}\sT$. Accordingly, particles moving along the direction of propagation of the GWs are deflected with an angle which changes with time. However, the order of magnitude of the displacement $\Delta L$ with respect to the unperturbed path is  $\displaystyle \Delta L \simeq \frac{S_{Y}A}{mc}$ (where $m$ is the mass of the particles), which is incredibly small. 

{We emphasize that our approach, which considers the interaction of gravitational waves  with the two spin degrees of freedom, can be applied to other analogous interactions with two-level systems where similar resonance conditions are present. Furthermore, the interaction Hamiltonian (\ref{eq:Hcoupling}) suggests that the spin degrees of freedom may in principle have an impact in interferometry experiments.}

In conclusion, we investigated the interaction between gravitational waves  and a quantum two-level system consisting of a spin 1/2  particle, highlighting the relevance of the gravitomagnetic field of the wave. We noted that resonance phenomena can occur regardless of the amplitude of the incident waves. Regarding the potential to leverage this effect for developing new detection techniques, we suggest focusing on the evolution induced by the passage of GWs through a sample with a large number of spins. This could, in turn, alter the macroscopic properties of the sample, such as its magnetization. This topic will be the subject of future research.

\section*{Acknowledgements}

The author expresses gratitude to the local research project "Modelli gravitazionali per lo studio dell'universo" (2022) from the Department of Mathematics "G. Peano," Università degli Studi di Torino, as well as to the Gruppo Nazionale per la Fisica Matematica (GNFM) for their contributions.

\appendix

\section{Derivation of the gravitoelectric and gravitomagnetic field of a plane monochromatic gravitational wave} \label{sec:app}

{Here, we show how to calculate the components of the gravitomagnetic field associated to a plane monochromatic gravitational wave. The metric (\ref{eq:weakfieldmetric11}) can be written in the form $g_{\mu\nu}=\eta_{\mu\nu}+h_{\mu\nu}$, where $h_{\mu\nu}$ is a perturbation of the flat Minkowski spacetime $\eta_{\mu\nu}$. Up to linear order in the perturbation $h_{\mu\nu}$, we can write the following expressions for the Riemann tensor \cite{MTW}:
\beq
R_{ikjl}=\frac 1 2 \left(h_{il,jk}+h_{kj,li}-h_{kl,ji}-h_{ij,lk} \right) \label{eq:riemann0}
\eeq
and
\beq
R_{ij0l}=\frac 1 2 \left( h_{il,j0}-h_{jl,i0} \right). \label{eq:riemann1}
\eeq
In particular, we consider the TT coordinates $(ct,x,y,z)$, so that the components for a wave propagating along the $x$ direction are \cite{Ruggiero:2021qnu}
\beq
h_{xx}=1, \quad h_{yy}=1-A^{+}\sin\left(\omega t-kx\right), \quad  h_{zz}=1+A^{+}\sin\left(\omega t-kx\right), \quad  h_{zy}=-A^{\times}\cos\left(\omega t-kx\right), \label{eq:perturb}
\eeq
where  $\omega$ is the frequency and $k$ the wave number, so that the wave four-vector is $\displaystyle k^{\mu}=\left(\frac \omega c, k, 0, 0 \right)$, with $k^{\mu}k_{\mu}=0$; $A^{+}, A^{\times}$ are the amplitudes of the wave in the two polarization states.
Utilizing gauge invariance in the linear approximation \cite{straumann2013applications}, we employ the above expressions to compute the Riemann tensor in Fermi coordinates: differently speaking, we can use the TT values for the perturbations $h_{\mu\nu}$ given by Eq.  (\ref{eq:perturb}) and express them in Fermi coordinates. It is important to note that in the metric (\ref{eq:weakfieldmetric11}), the Riemann tensor is evaluated along the reference world-line. Consequently, after deriving the components of the Riemann tensor using Eq. (\ref{eq:perturb}), we set $X=0$. In other words, we suppose that the extension of the reference frame is much smaller than the wavelength, so that we may neglect the spatial variation of the wave field: if this condition is not fulfilled, it is necessary to use the expression of the Fermi coordinates valid at higher order in the distance from the reference world-line, as discussed by \citet{marzlin}.\\
\indent In particular, the gravitoelectric potential (\ref{eq:defPhiG}) is
\beq
\Phi=\frac{\omega^{2}}{4}\left[ A^{+}\sT Y^{2}+2A^{\times}\cT YZ-A^{+}\sT Z^{2} \right], \label{eq:defPhicomp}
\eeq
while the components of the gravitomagnetic potential (\ref{eq:defAG})  are
\begin{eqnarray}
\mathcal{A}_{X}&=&\frac{\omega^{2}}{6} \left[A^{+}\sT (Y^{2}-Z^{2})+2A^{\times}\cT ZY \right], \label{eq:defAX} \\
\mathcal{A}_{Y}&=& \frac{\omega^{2}}{6} \left[-A^{+}\sT YX-A^{\times}\cT ZX \right], \label{eq:defAY} \\
\mathcal{A}_{Z}&=& \frac{\omega^{2}}{6} \left[-A^{\times}\cT YX+A^{+}\sT XZ \right].  \label{eq:defAZ}
\end{eqnarray}
From the above expressions, we obtain the components of the gravitoelectric field:
\small
\beq
E^{}_{X}  = 0, \quad E^{}_{Y}  = -\frac{\omega^{2}}{2}\left[A^{+} \sin \left(\omega T \right)Y+A^{\times} \cos \left(\omega T \right) Z \right], \quad E^{}_{Z}  = -\frac{\omega^{2}}{2}\left[A^{\times}\cT Y-A^{+}\sT Z \right], \label{eq:campoE}
\eeq
\normalsize
and those of the gravitomagnetic field:
\small
\beq
B^{}_{X}  = 0, \quad B^{}_{Y}  = -\frac{\omega^{2}}{2}\left[-A^{\times} \cT Y+A^{+} \sT Z \right], \quad B^{}_{Z}  = -\frac{\omega^{2}}{2}\left[A^{+}\sT Y+A^{\times}\cT Z \right]. \label{eq:campoB}
\eeq}
\normalsize

\bibliography{quantum_GW}

\end{document}